\documentclass[letterpaper, 10 pt, conference]{ieeeconf}  
\IEEEoverridecommandlockouts           
\usepackage{cite}
\usepackage{amsmath,amssymb,amsfonts}
\usepackage{graphicx}
\usepackage{textcomp}
\usepackage{xcolor}
\def\BibTeX{{\rm B\kern-.05em{\sc i\kern-.025em b}\kern-.08em
    T\kern-.1667em\lower.7ex\hbox{E}\kern-.125emX}}

\usepackage{comment}
\usepackage[utf8]{inputenc}
\usepackage{amsmath,graphicx}
\usepackage{mathrsfs}
\usepackage{xcolor}
\usepackage{mathtools}
\usepackage{enumerate}
\usepackage{booktabs}
\usepackage{romannum}
\usepackage{float}
\usepackage{algorithm}
\usepackage{dirtytalk}

\usepackage{subcaption}

\newcommand{\mbu}{\mathbf{u}}
\newcommand{\mbv}{\mathbf{v}}

\newcommand{\mbx}{\mathbf{x}}

\newcommand{\mbz}{\mathbf{z}}

\newcommand{\mbe}{\mathbf{e}}

\newcommand{\cA}{\mathcal{A}}
\newcommand{\cB}{\mathcal{B}}

\newcommand{\cI}{\mathcal{I}}

\newcommand{\cL}{\mathcal{L}}

\newcommand{\cR}{\mathcal{R}}
\newcommand{\cS}{\mathcal{S}}

\newcommand{\cU}{\mathcal{U}}

\newcommand{\bR}{\mathbb{R}}

\newcommand{\bpsi}{\boldsymbol{\psi}}

\renewcommand{\baselinestretch}{0.998}

\usepackage{algpseudocode}

\usepackage{amsthm}
\newtheorem{theorem}{Theorem}
\newtheorem{lm}{Lemma}
\newtheorem{corr}{Corollary}

\newtheorem{remark}{Remark}
\theoremstyle{definition}

\newtheorem{problem}{Problem}

\usepackage[normalem]{ulem}

\newcommand{\x}[1]{\mathbf{x}_{#1}}

\newcommand{\ball}[2]{\mathcal{B}_{#1}\left( #2 \right)}

\newcommand{\distance}[2]{\|#1-#2\|}
\newcommand{\unitVector}[2]{\mbe(#1,#2)}

\makeatletter
\renewcommand*\env@matrix[1][\arraystretch]{%
  \edef\arraystretch{#1}%
  \hskip -\arraycolsep
  \let\@ifnextchar\new@ifnextchar
  \array{*\c@MaxMatrixCols c}}
\makeatother

\newcommand{\RR}{{\mathbb{R}}}

\DeclareUnicodeCharacter{2212}{-}
\title{\Large\bf 
Heterogeneous Roles against Assignment Based Policies in Two vs Two Target Defense Game
}
\author{Goutam Das$^{1}$, 
Violetta Rostobaya$^{1}$,
James Berneburg$^{1}$,
Zachary I. Bell$^{2}$,
Michael Dorothy$^{3}$,  and 
Daigo Shishika$^{1}$
\thanks{We gratefully acknowledge the support of  ARL grant ARL DCIST CRA
W911NF-17-2-0181 and AFRL grant FA8651-23-1-0012.
The views expressed in this paper are those of the authors and do not reflect the official policy or position of the United States Government, Department of Defense, or its components.}
\thanks{$^{1}$Goutam Das, Violetta Rostobaya, James Berneburg, and Daigo Shishika are with George Mason University, 
4400 University Dr, Fairfax, VA 22030, USA {\tt\small \{gdas,vrostoba,jbernebu,dshishik\}@gmu.edu}}%
\thanks{$^{2}$Zachary I. Bell is with the Munitions Directorate, Air Force Research Laboratory, Eglin AFB, FL 32542, USA {\tt\small zachary.bell.10@us.af.mil}}%
\thanks{$^{3}$Michael Dorothy. Army Research Directorate, DEVCOM Army Research Laboratory, APG, MD.
         {\tt\small michael.r.dorothy.civ@army.mil}}%
}
\begin{document}
\maketitle 
\thispagestyle{empty}
\pagestyle{empty}

\begin{abstract}
In this paper, we consider a target defense game in which the attacker team seeks to reach a high-value target while the defender team seeks to prevent that by capturing them away from the target. To address the curse of dimensionality, a popular approach to solve such team-vs-team game is to decompose it into a set of one-vs-one games. Such an approximation assumes independence between teammates assigned to different one-vs-one games, ignoring the possibility of a richer set of cooperative behaviors, ultimately leading to suboptimality.
In this paper, we provide teammate-aware strategies for the attacker team and show that they can outperform the assignment-based strategy, if the defenders still employ an assignment-based strategy. 
More specifically, the attacker strategy involves heterogeneous roles where one attacker actively intercepts a defender to help its teammate reach the target.
We provide sufficient conditions under which such a strategy benefits the attackers, and we validate the results using numerical simulations.
\end{abstract}

\section{Introduction}
\noindent Differential game theory provides a set of mathematical tools particularly useful for team-vs-team engagements in multi-agent scenarios, stands out for their strategic depth and applicability across diverse fields such as robotics, security, and military strategy \cite{weintraub2020introduction}. 
Isaacs first introduced these tools in his seminal work \cite{Issacs1965} in which he formulated and solved a class of problems known as target-defense pursuit-evasion games. These games, characterized by the dynamic interplay between pursuers and their targets, highlighted an intricate balance between collective objectives and individual roles.

Subsequently, several variations of the problem have been introduced, including Perimeter-defense \cite{von2022circular, das2022guarding, bajaj2022competitive}, Border-defense \cite{lee2024Guarding,garcia2019strategies, Garcia2021_NvM}, and Reach-avoid games \cite{bakolas2010optimal, pachter2022strategies, dorothy2024one, goutam2024defending}. When the number of agents involved is large, obtaining the exact solutions become intractable due to their high dimensionality. Both geometric and analytical approaches have been considered to provide performance bounds. Particularly in multi-agent games, a common approach to make the problem tractable is to approximate the complex team interactions into a combination of one-vs-one (1v1) engagements\cite{Kuhn1955Hungarian,Shishika2018,sun2019multiplayer,yan2019task}. While such decomposition method offers analytical tractability, it overlooks the potential cooperative behavior among the attacker team members. 

 In \cite{Garcia2021_NvM, bakolas2021decentralized, zhang2021pursuer, mittal2018pursuit}, complex team-vs-team games are analyzed, where two teams compete against each other to maximize their payoff. In all these scenarios, the optimal solution was derived using the decomposition approach. Inspired by these works, we study a special case of team-vs-team target-defense game (TT-TDG) in which a team of two defenders seeks to capture an equal number of attacker agents before they can reach a target point. We consider defenders that adopt the traditional assignment-based approach, which can be proved optimal provided that different defender-attacker pairs do not interact with each other \cite{Garcia2021_NvM}. 
 We are interested in a case where this independence assumption does not hold: i.e., when one attacker takes a role to help its teammate instead of trying to win in a greedy manner. 
 We demonstrate that such a teammate-aware strategy allows the attacker team to improve its performance under certain game conditions.

\subsubsection*{Statement of contributions} The main contributions of this work are:
(i) identifying and illustrating the limitations inherent in decomposing team-vs-team games into individual 1v1 engagements; 
(ii) highlighting the importance of cooperative strategies in team-vs-team games to improve overall team performance; 
(iii) developing a set of team-aware feedback strategies for the attacker team that can effectively counter the defenders' assignment-based approaches; and 
(iv) identifying a set of initial states from which the attacker team can feasibly adopt these alternative strategies.

\subsubsection*{Organization of the paper} The paper is organized into the following sections: Section II formulates the 2v2-TDG. Section III discusses the existing solution approaches for 2v2-TDG. Section IV presents the main results on attackers' heterogeneous role selection and deviation from the existing solution with performance bounds. Section V shows three examples to demonstrate the results. Finally, Section VI provides concluding remarks and directions for future research.

\section{Problem Formulation}
\subsection{Notations}
\noindent We denote by $\|\cdot\|$ the Euclidean norm of a vector. Let $\mathbb{R}$, $\mathbb{R}^+$, and $\mathbb{Z}^+$ represent the set of real numbers, positive real numbers, and positive integers, respectively. 
Let $\ball{\rho}{\mathbf{x}}$ and $\partial\ball{\rho}{\mathbf{x}}$ denote the closed ball and its boundary, respectively, both with radius $\rho \in \mathbb{R}^+$ centered at the point $\mathbf{x} \in \mathbb{R}^2$.
The line segment joining points $\mathbf{x}_1$ and $\mathbf{x}_2$ in $\mathbb{R}^2$ is denoted by $\cL(\mathbf{x}_1,\mathbf{x}_2)$.
The unit vector from $A$ to $B$ is denoted as 
\begin{equation}
\mathbf{e}({\mbx_A,\mbx_B}) \triangleq \frac{\mathbf{x}_B - \mathbf{x}_A}{\left\|\mathbf{x}_B - \mathbf{x}_A\right\|}.
\end{equation}

\subsection{Problem Statement}
\noindent We consider a special case of team-vs-team target-defense game (TT-TDG) on a plane played between two attackers ($A_i$) and two defenders ($D_j$), and a stationary target ($T$), where $i,j \in \cI = \{1,2\}$ are the indices of the agents. The positions of the agents are
denoted as $\mbx_k = [x_k, y_k]^\top$, where $k \in \{T, A_i, D_j\}$. Without loss of generality, we consider $T$ to be at the origin, i.e., $\mbx_T = [0, 0]^\top$. The agents' kinematics are as follows:
\begin{align}\label{eq: kinematics}
&\dot{\x{}}_{A_i}(t) = \nu \mbu_i(t), \\ 
    &\dot{\x{}}_{D_j}(t) = \mbv_j(t),
\end{align}
where $\mbu_i, \mbv_j \in \mathcal{U}$ are the $i$th attacker's and $j$th defender's inputs and $\nu \in (0, 1)$ is the speed ratio.
We use $\mathcal{U}$ 
to denote the set of admissible controls of attackers and defenders, where 
\begin{align} \label{eq: mathcal_U}
    \mathcal{U} \triangleq \{\x{} \in \RR^2 \mid \|\x{}\| \leq 1 \}.
\end{align}

We assume that the players have the perfect knowledge of $\mbz \triangleq \left[ \x{D_1}^\top, \x{D_2}^\top, \x{A_1}^\top, \x{A_2}^\top \right]^\top \in \RR^{8}$ at any given time $t$. We use $\gamma_{A_i},\gamma_{D_j} \in \Gamma$, where $\Gamma = \{\gamma :\bR^{8} \to \cU\}$, to denote the state feedback strategies of the attackers and defenders, respectively. 
We evaluate these policies at $\mathbf{z}$ to obtain the following concatenated team strategies in ordered pairs:
\begin{subequations}
\begin{align}
\mathbf{u}(\mathbf{z}) = (\gamma_{A_1}(\mathbf{z}), \gamma_{A_2} (\mathbf{z})) \in \Gamma^2,  
\\
\mathbf{v}(\mathbf{z}) = (\gamma_{D_1}(\mathbf{z}), \gamma_{D_2}(\mathbf{z})) \in \Gamma^2.
\end{align}
\end{subequations}
The game ends when: a) the target is \emph{captured} by an attacker (attackers win scenario), or b) every attacker is captured by a unique defender (defender win scenario).
We assume that, when defender $j$ captures attacker $i$, both are taken out of play, and they can no longer capture or be captured.
This means that, for a defender win, the game must end when each attacker is captured by a \emph{unique} defender. 
To formalize this notion, let $\Psi_\mathcal{I} \subset \mathcal{I}^2$ denote the set of all permutations of $\mathcal{I}$. 
Formally the terminal manifold, $\cS$ is defined as follows:
\begin{align} \label{eq: mathcal_S term man.}
    \cS = \cS_A \cup \cS_D,
\end{align}
where
\begin{subequations}
    \begin{align}
        \mathcal{S}_A &= \left\{ \mathbf{z} \mid \exists i \in \mathcal{I}, \distance{\mathbf{x}_{A_i}}{\mathbf{x}_T} = 0 \right\}, \label{eq: mathcal_SA A win term man} \\ 
        \mathcal{S}_D &= \left\{ \mathbf{z} \mid  \exists \bpsi' \in \Psi_\mathcal{I} \text{ s. t. }\mathbf{x}_{A_i} = \mathbf{x}_{D_{\bpsi_i'}}, \forall i \in \mathcal{I}  \right\}. \label{eq: mathcal_SD D win term man}
    \end{align}
\end{subequations}
Here $\mathcal{S}_A$ and $\mathcal{S}_D$ represents terminal manifolds for attacker and defender win, respectively. 
The final time $t_f$ corresponds to the first time instant at which $\mbz$ enters one of the terminal manifolds, i.e.,
$t_f=$ $\inf \left\{t \in \mathbb{R}_{\geq 0}: \mbz(t) \in \cS_A \cup \cS_D \right\}$.
We use $t_f^{D_j}$ and $t_f^{A_i}$ to denote the final time that each agent is active. Given the initial condition  $\mbz(0) = \mbz_0$, the (terminal) payoff of this game is then defined as:
\begin{align}\label{eq: J payoff}
 J(\mathbf{u}(\mathbf{\cdot}), \mathbf{v}(\mathbf{\cdot}); \mathbf{z}_0) = \min_{i \in \mathcal{I}} \distance{\mathbf{x}_{A_i}(t_f^{A_i})} {\mathbf{x}_T}.   
\end{align}
The objective of the attacker (resp. defender) team is to collectively minimize (resp. maximize) $J$. 
The Value of the game, if it exists, is given by 
\begin{equation} \label{eq: V value}
V\left(\mathbf{z}_0\right)=\min _{\mbu \in \Gamma^2} \max _{\mbv \in \Gamma^2} J\left(\cdot ; \mathbf{z}_0\right)=\max _{\mbv \in \Gamma^2} \min _{\mbu \in \Gamma^2} J\left(\cdot ; \mathbf{z}_0\right).
\end{equation}

The payoff functional given in equation \eqref{eq: J payoff} illustrates a \emph{game of degree}, quantifying the margin of win at $t_f$. In contrast, a \emph{game of kind} classifies outcomes into finite categories, such as an attacker win (if $J=0$ ) or a defender win (if $J>0$ ).

\paragraph*{Gap in the Literature}
Given a target-defense game between and team of defenders and attackers on a plane described by kinematics \eqref{eq: kinematics},  terminal manifold \eqref{eq: mathcal_S term man.} and payoff \eqref{eq: J payoff}, we address a gap in traditional solution approaches. Literature often simplifies team versus team engagements into a set of one-versus-one (1v1) games, aggregating individual outcomes to deduce the overall result. This method overlooks the interaction between teammates, where:
\begin{itemize}
    \item Each defender $D_j$ is assigned to an attacker $A_i$, leading to two distinct 1v1 engagements.
    \item Upon fixing defender assignments, any deviation by attackers from their prescribed state-feedback strategies inherently advantages the defenders\cite[Remark 1]{Garcia2021_NvM}.
\end{itemize}
This raises pivotal questions regarding the attackers' strategic considerations and the conditions under which deviations from the \emph{nominal} strategies could enhance their collective outcome, thereby challenging the defenders' anticipations. Specifically, we seek to address the following problems:


\begin{problem}
Identify the initial conditions under which attackers, by embracing individual roles within their team and possibly deviating from state-feedback nominal strategies, could effectively alter the game's outcome in their favor.
\end{problem}

\begin{problem}
Determine the resultant dynamics and outcomes when attackers exploit such strategic deviations, potentially leading to a more advantageous collective outcome than what is predicted by analyzing aggregated 1v1 engagements.
\end{problem}

\section{Existing Solution} 
\begin{figure}[bp]
    \includegraphics[width=0.49\textwidth]{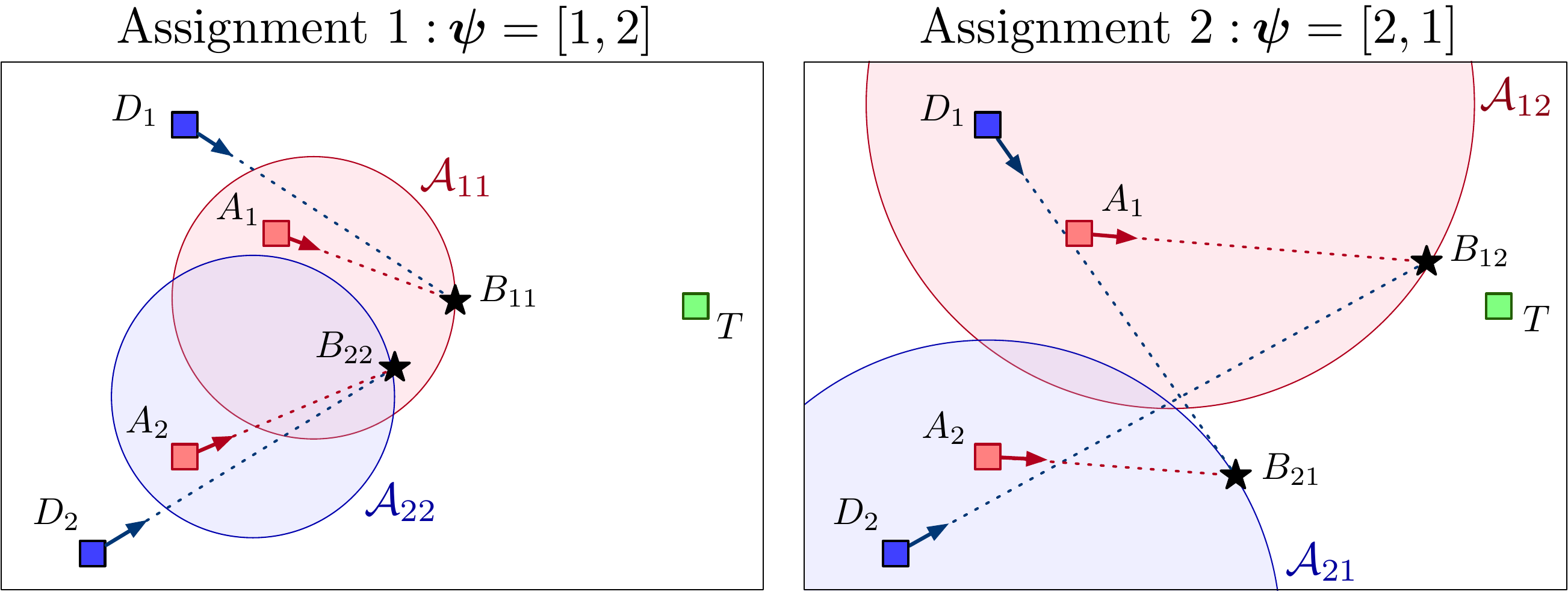}
    \caption{Example of 2v2 game with two possible defender-attacker assignments, both of which resulting in defender's win but with different payoff. The assignment 1 is preferred by the defender team. 
    }
    \label{fig: assignments}
\end{figure}

\noindent In this section we discuss the traditional approach of solving 2v2-TDG as a two-step optimization problem illustrated in Fig.~\ref{fig: assignments}: \textbf{(a)} optimal cost $\phi_{ij}$ for each assigned defender ($D_j$) to the attacker ($A_i$), and \textbf{(b)} optimal assignment ($\bpsi(i)$) of the defenders that maximizes $J$.
The first problem is usually solved using geometric method (Apollonius Circle (AC)), and the second problem requires Linear-Bottleneck Assignment policy (LBAP), which we are going to discuss next.

\subsection{Solution of 1v1 Game}\label{subsection: dominance region}
\noindent For a target-defense game, let $\mbx_{ij} = [\mbx_{A_i}^\top, \mbx_{D_j}^\top]$ denote the conjoint state. The AC defined by this pair of agents $\cA_{ij}$ is defined as follows:
\begin{equation}\label{eq: AC def.}
    \mathcal{A}_{ij}(t) =\left\{\mbx \in \bR^2: \distance{\mbx}{\mbx_{C_{ij}}} \leq \rho_{ij}\right\},
\end{equation}
where its center $\x{C_{ij}}$ and radius $\rho_{ij}$ are given by: 
\begin{equation}\label{eq: xC}
    \begin{aligned}
        \x{C_{ij}} =  \alpha \x{A_i}- \beta\x{D_j}, \text{ and } \rho_{ij} = \gamma \distance{\mbx_{A_i}}{\mbx_{D_j}},
    \end{aligned}
\end{equation}
with $\alpha = \frac{1}{1-\nu^2}$, $\beta = \frac{\nu^2}{1-\nu^2}$, and $\gamma = \frac{\nu}{1-\nu^2}$. 
Intuitively, because the attacker is slower, the interior of the AC is the set of points which the attacker can reach before the defender does, and vice versa for the points outside the circle. 

For a single defender and a single attacker \eqref{eq: J payoff} becomes
\begin{align}\label{eq: 1v1 J payoff}
    J^{1v1}(\gamma_{A_i},\gamma_{D_j}; \mbx_{ij}(0)) = \distance{\mbx_{A_{i}}(t_f^{A_{i}})}{\mbx_T}.
\end{align}

Let $\x{B_{ij}}$ be the optimal \emph{capture point},
which is the point on $\cA_{ij}$ closest to the target: 
\begin{equation}\label{eq: xB 1v1}
        \x{B_{ij}} = \text{arg} \min_{\x{}\in \cA_{ij}} \distance{\mbx}{\mbx_T}.
\end{equation}

Note that the headings of $A_i$ and $D_j$ must be constant under optimal play, and the optimal trajectories are straight lines. This is true when the agents have single integrator dynamics and the terminal payoff is Mayer type \cite{Garcia2021_NvM}.
The saddle-point equilibrium feedback-strategies for the players can be derived as follows \cite{Garcia2021_NvM}:
\begin{subequations}
    \begin{align} 
    \gamma_{A_i}^*(\mbx_{ij}) = \unitVector{\mbx_{A_i}}{\mbx_{B_{ij}}}, \label{eq: gamma_A 1v1} \\
   \gamma_{D_j}^*(\mbx_{ij}) = 
   \unitVector{\mbx_{D_j}}{\mbx_{B_{ij}}}, \label{eq: gamma_D 1v1}
    \end{align}
\end{subequations}
which suggests that under the optimal play, both players move towards the current optimal capture point. 
Using $\mbx_{B_{ij}}$, we can define the equilibrium outcome of this 1v1 game~\eqref{eq: 1v1 J payoff} as follows:
\begin{align}
   \phi_{i j} \triangleq
   \begin{cases}
       \distance{\mbx_{B_{ij}}}{\mbx_T}, &\text{if } \mbx_T \notin \mathcal{A}_{ij},
       \\
       0, &\text{otherwise.}
   \end{cases}
\end{align}

\subsection{Linear Bottleneck Assignment Policy (LBAP)}
\noindent We can write the assignment strategy of the defenders as a permutation of $\mathcal{I}$, represented by a vector $\bpsi \in \Psi_\mathcal{I}$, where an element $\bpsi_i\in \mathcal{I}$ indicates the index of the defender that is assigned to attacker $i$. Given the payoff function~\eqref{eq: J payoff}, the assignment of the defenders to the attackers can be found by linear bottleneck assignment policy \cite{burkard2012assignment}, which can be cast as the following linear program: 
\begin{align}\label{eq: lbap}
      \bpsi^* = \text{arg} \max_{\bpsi \in \Psi_\mathcal{I}} \min_{i \in  \mathcal{I}}  \phi_{i \bpsi_i}.
\end{align}
Based on the optimal assignment we propose two definitions.

\textbf{Definition 1 (Critical Attacker-Defender Pair):} 
We define the critical attacker-defender pair $(A_{i^*}, D_{j^*})$ as the pair that corresponds to the minimum cost in the optimal assignment $\bpsi^*$:
\begin{equation}
i^* = \underset{i \in \mathcal{I}}{\arg\min} \; \phi_{i \bpsi^*_i},
\end{equation}
where $j^* = \bpsi^*_{i}$ denotes the defender $D_{j^*}$ assigned to the critical attacker $A_{i^*}$, minimizing the payoff function $\phi_{ij}$.

Non-critical attacker-defender pairs are defined as all pairs $(A_i, D_j)$, where $i \neq i^*$ and $j = \bpsi^*_i$, indicating that these pairs do not correspond to the minimum cost in the optimal assignment $\bpsi^*$.

\subsection{Nominal Strategies in 2v2 Game}\label{subsec: nom 2v2}
\noindent For simplicity, we assign critical attacker and defender indices to be $i^*=j^*=1$, and for non-critical attacker and defender $i=j=2$.
From \eqref{eq: lbap} along with with the Definition~1, the optimal assignment from~\eqref{eq: lbap} is $\bpsi^* = [1,2]^\top$ and subsequently, optimal cost $\phi^* \triangleq \phi_{11} \leq \phi_{22}$.

Additionally, in this section onward we restrict our analysis to the defender-winning case, i.e., $\x{T}\notin \mathcal{A}_{11}(0)\cup \mathcal{A}_{22}(0)$. Once the assignment is determined by the defenders, the \emph{nominal} strategies for the critical attacker, $A_1$, and the critical defender, $D_1$, are:
\begin{subequations}
    \begin{align}
    \gamma_{A_1}^*(\mbz_0)
    =  \unitVector{\mbx_{A_1}}{\mbx_{B_{11}}}, \label{eq: nom strategies A1} \\   
    \gamma_{D_1}^*(\mbz_0)
    = \unitVector{\mbx_{D_1}}{\mbx_{B_{11}}}\label{eq: nom strategies D1}. 
    \end{align}
\end{subequations}
Similarly, the nominal strategy for non-critical players, $A_2$ and $D_2$, are:
\begin{subequations}
    \begin{align}
    \gamma_{A_2}^*(\mbz_0)
    =  \unitVector{\mbx_{A_2}}{\mbx_{B_{22}}} \label{eq: nom strategies A2},   
\\
    \gamma_{D_2}^*(\mbz_0)
    = \unitVector{\mbx_{D_2}}{\mbx_{B_{22}}} \label{eq: nom strategies D2}.  
    \end{align}
\end{subequations}
From Lemma~\ref{lemma: xB_dot}, if all players use their respective nominal strategies, capture points, $\x{B_{11}}$ and $\x{B_{22}}$, remain stationary.
For brevity, we denote $\mathbf{x}_B \triangleq \mathbf{x}_{B_{11}}$.
The nominal payoff of the 2v2 game is:
\begin{equation}\label{eq:nominal payoff}
   \Phi(\mbz_0) \triangleq J(\mbu^*(\cdot),\mbv^*(\cdot);\mbz_0)= \distance{\mbx_{B}}{\mbx_T}.
\end{equation}

Once the optimal assignment $\bpsi^*$ is determined, there is an implicit assumption that interception occurs only with the assigned defenders. This assumption ensures that following the strategies described by \eqref{eq: nom strategies A1} and \eqref{eq: nom strategies A2} is necessary and sufficient to guarantee the payoff $\Phi$.

However, this framework does not account for potential cooperative strategies among attackers that could disrupt the defenders' assignments and improve the attackers' payoffs. In the subsequent section, we will explore the attackers' strategy space, examining whether the defenders' assignment-based approach to determining state-feedback equilibrium strategies is sufficient to secure the intended outcomes in 2v2-TDG.

\section{Main Results}
\noindent In this section, we explore scenarios in which the attacker team benefits from deviating from the nominal strategies described previously. Our analysis focuses on identifying initial conditions that allow cooperative behaviors among attacker agents, enabling them to adopt heterogeneous roles that contribute to the team's performance. We further discuss two distinct scenarios wherein deviations from these nominal strategies—by one or both attackers—yield an improved payoff for the attackers. 

 First, we decompose the game into two potential phases, delineated by the sequence of captures. The terminal time of Phase~I is $t_{f_1} \leq t_f$, where $t_{f_1}$ is the time of the first capture, and $t_f$ is the terminal time of Phase~II and the game. It is possible for the first capture to occur when the game ends, such as when simultaneous capture occurs, in which case Phase II has zero time duration. 

  Note that Phase~I is the only phase in which the attackers can possibly have the incentive to deviate from the nominal strategies. 
  This is because Phase~I is the only team-vs-team phase for the 2v2 case studied in this paper, and Phase~II is a 1v1 scenario for which equilibrium solution is known and is given by \eqref{eq: gamma_A 1v1}.
Therefore, our discussion primarily focuses on Phase~I. 
\subsection{One Attacker Deviation}\label{sec:attackerDeviates1}
\begin{figure}[bp]
    \centering
    \includegraphics[width=0.49\textwidth]{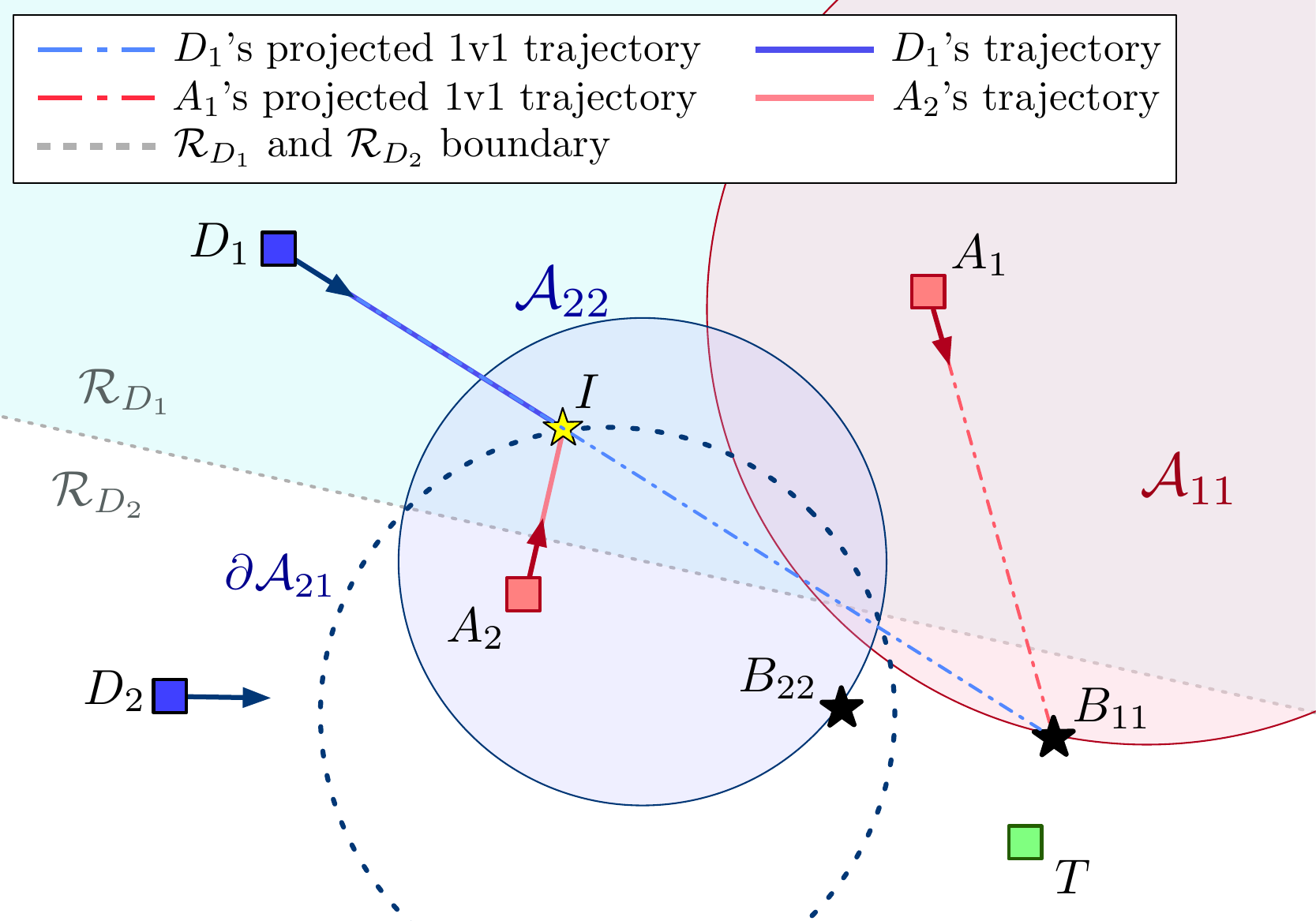}
    \caption{Illustration of $A_2$'s deviation from the nominal strategy to intercept $D_1$ at $\mbx_I$.
    }
    \label{fig: one_att_dev}
\end{figure}
\noindent We first look into the scenario where only the non-critical attacker, $A_2$ deviates from \eqref{eq: nom strategies A2}, which is illustrated in Fig.~\ref{fig: one_att_dev}. We assume that all other agents follow the nominal strategies described in Section~\ref{subsec: nom 2v2}. 
\begin{lm} \label{lemma: xB_dot}
The time derivative of the optimal capture point,  $\mbx_{B} \triangleq \mbx_{B_{ij}}$ can be written as follows:
\begin{equation}\label{eq: xB_dot}
    \dot{\mathbf{x}}_B = 
\gamma\frac{\mathbf{a}^\top\dot{\mathbf{a}}}{\|\mathbf{a}\|}\cdot\frac{\mathbf{b}}{\|\mathbf{b}\|}
        - \gamma\|\mathbf{a}\|\left(\frac{\mathbf{b}(\mathbf{b}^\top\dot{\mathbf{b}})}{\|\mathbf{b}\|^3}-\frac{\dot{\mathbf{b}}}{\|\mathbf{b}\|}\right),
\end{equation}
where $\mathbf{a}, \mathbf{b} \in \mathbb{R}^2$ are vectors defined as $\mathbf{a} =\mathbf{x}_{A_i}-\mathbf{x}_{D_j}$, and $\mathbf{b} =\mathbf{x}_T-\mathbf{x}_{C_{ij}}$, respectively.
 If both the attacker and the defender employ their respective state-feedback strategies from \eqref{eq: gamma_A 1v1} and \eqref{eq: gamma_D 1v1}, then $\dot{\mathbf{x}}_B = 0$; if either of them deviate, then $\dot{\mathbf{x}}_B \neq 0$.
\end{lm}

\begin{proof}
The optimal capture point $\mathbf{x}_{B_{ij}}$ can be expressed as
\begin{equation} \label{eq: xB_explicit}
    \mathbf{x}_{B_{ij}} = \mathbf{x}_{C_{ij}} + \rho_{ij} \unitVector{\mathbf{x}_{C_{ij}}}{\mathbf{x}_T}.
\end{equation}
Differentiating \eqref{eq: xB_explicit} yields \eqref{eq: xB_dot}. By substituting the controls for the attacker and defender, it 
can be shown that $\dot{\mathbf{x}}_B = 0$ under equilibrium conditions, and $\dot{\mathbf{x}}_B \neq 0$ for any unilateral deviation from the equilibrium strategies.
\end{proof}
\begin{remark}\label{remark: straightline-condition}
    From Lemma~\ref{lemma: xB_dot} if defender follows \eqref{eq: gamma_D 1v1}, the defender's trajectory remains a straight line iff the corresponding attacker follows \eqref{eq: gamma_A 1v1}.
\end{remark}

From Remark~\ref{remark: straightline-condition}, the trajectory of $D_1$ is a straight line. However, since $A_2$ deviates from the nominal strategy, the trajectory of $D_2$ may not be straight. This deviation raises the question of which defender reaches a certain point first. 
Given that both $D_1$ and $D_2$ have equal speed, we define the set of points that $D_1$ will reach before $D_2$ does as follows:
\begin{equation} \label{eq: R_D1}
    \cR_{D_1} =  \left\{\mathbf{x} \in \mathbb{R}^2 \mid \distance{\mathbf{x}}{\mathbf{x}_{D_1}} < \distance{\mathbf{x}}{\mathbf{x}_{D_2}} \right\}.
\end{equation} 
Similarly, we can define $\cR_{D_2}$, the set of points that $D_2$ can reach before $D_1$.

Now we are ready to present the initial condition for $A_2$'s deviation in the following theorem.

\begin{theorem}[Condition for Non-Critical Attacker Deviation] \label{theorem: 1. one-dev cond.}
Given a 2v2-TDG with
dynamics \eqref{eq: kinematics}, payoff \eqref{eq: J payoff}, and the state-feedback strategies of $A_1$, $D_1$, and $D_2$ given in \eqref{eq: nom strategies A1}, \eqref{eq: nom strategies D1}, and  \eqref{eq: nom strategies D2}, respectively, $A_2$ can deviate from \eqref{eq: nom strategies A2} and terminate Phase~I with 
${\mbx_{D_1}(t_{f_1})}={\mbx_{A_2}(t_{f_1})},$
if 
\begin{equation} \label{eq: intercept points xI}
\exists \mathbf{x}_I \in \mathbb{R}^2 \text{ s. t. } \mathbf{x}_I \in \cR_{D_1} \cap \mathcal{L}(\mathbf{x}_{D_1}, \mathbf{x}_B) \cap \mathcal{A}_{21}(0), 
\end{equation}
where $\cA_{21}(0)$ is the AC of $D_1$ and $A_2$ defined in \eqref{eq: AC def.} at initial time $t=0$.
\end{theorem}
\begin{proof}
The proof follows from the fact that any unilateral deviation in Phase~I by $A_2$ does not alter the optimal assignment of the defenders, because in equilibrium $\distance{\mbx_{A_{1}}(t_f)}{\mbx_T} \leq \distance{\mbx_{A_{2}}(t_f)}{\mbx_T}$, i.e., $A_2$ can never become the critical one unless $A_1$ deviates.
Furthermore, given that $\mbx_I \in \cR_{D_1} \cap \cA_{21}$, it is guranteed that $A_2$ can reach $\mbx_I$ prior to any potential capture by $D_2$. This provides $A_2$ the opportunity to choose $\mbx_I = \mbx_{D_1}(t_{f_1})$, as an \emph{interception point}, to alter the outcome of the game.
\end{proof}
\begin{remark}\label{rem: sacrificial A2}
Notice that Theorem~\ref{theorem: 1. one-dev cond.} provides a sufficient condition for the non-critical attacker $A_2$ to take a sacrificial role, thereby eliminating the critical defender $D_1$ from the game, assuming all other agents follows their nominal state-feedback strategies. This strategy effectively leaves the non-critical defender $D_2$ as the only pursuer of the critical attacker $A_1$ during Phase~II, for $t_2 \in (t_{f_1}, t_f]$.
\end{remark}

\begin{corr}[Straight-line Interception Strategy for $A_2$]\label{cor: 1}
Given a point $\mbx_I \in \partial \cA_{21}(0)$ which satisfies \eqref{eq: intercept points xI}, $A_2$ can intercept $D_1$ with a straight-line trajectory following a state-feedback strategy given as follows: 
\begin{align}\label{eq: A2 sac gamma one dev}
    \gamma_{A_2}(\mbz_0) = \unitVector{\mbx_{A_2}}{\mbx_I}.
\end{align}
\end{corr}
Note that $\mathbf{x}_I$ could represent more than one point, particularly when the nominal trajectory of $D_1$: $\cL(\mbx_{D_1},\mbx_{B})$, intersects $\partial \mathcal{A}_{21}$. It is evident that $A_2$ can reach both of these points and intercept $D_1$, however, the existence of at least one point is sufficient for the purpose of this paper, and identifying the optimal one is left for future work. 

\begin{lm} \label{lem: J improve after intercept}
If the initial conditions satisfy Corollary~\ref{cor: 1} and if $A_2$ follows the state-feedback strategy given by \eqref{eq: A2 sac gamma one dev}, then the attacker team can improve its payoff. 
Specifically, the payoff of the game is given by
\begin{equation}
 \begin{aligned}
        J = \phi_{12}(t_{f_1}) = \distance{\mbx_{B_{12}}(t_{f_1})}{\mbx_T},
    \\
    \text{and } \phi_{12}(t_{f_1})< \Phi,
 \end{aligned}
\end{equation}
where $\Phi$ is the nominal payoff given in \eqref{eq:nominal payoff}. Furthermore, if $\phi_{12}(t_{f_1}) = 0$, the attacker team can win the game.
\end{lm}
\begin{proof}
After $A_2$ strategically deviates to intercept $D_1$, as using \eqref{eq: A2 sac gamma one dev}, the outcome of the game changes, since $D_2$ must capture $A_1$. This switching in assignment ensures that $J$ reduces, based on the optimality of the original LBAP assignment. More specifically, $\phi_{12} \leq \phi_{11} = \Phi$. Furthermore, since $D_2$ did not play optimally against $A_1$ in Phase I. We also have $\phi_{12}(t_{f_1}) < \phi_{12}(0)$, which provides $\phi_{12}(t_{f_{1}}) < \Phi$. 
\end{proof}

Notice that, since the critical attacker $A_1$ adheres to the nominal feedback strategies at each phase of the game, its path is a piece-wise linear trajectory. 
Next, we will investigate scenarios in which $A_1$ also deviates from the nominal path adopting a straight-line path for the entire game.

\subsection{Two Attacker Deviation}
\begin{figure}[tbp]
\centering    \includegraphics[width=0.49\textwidth]{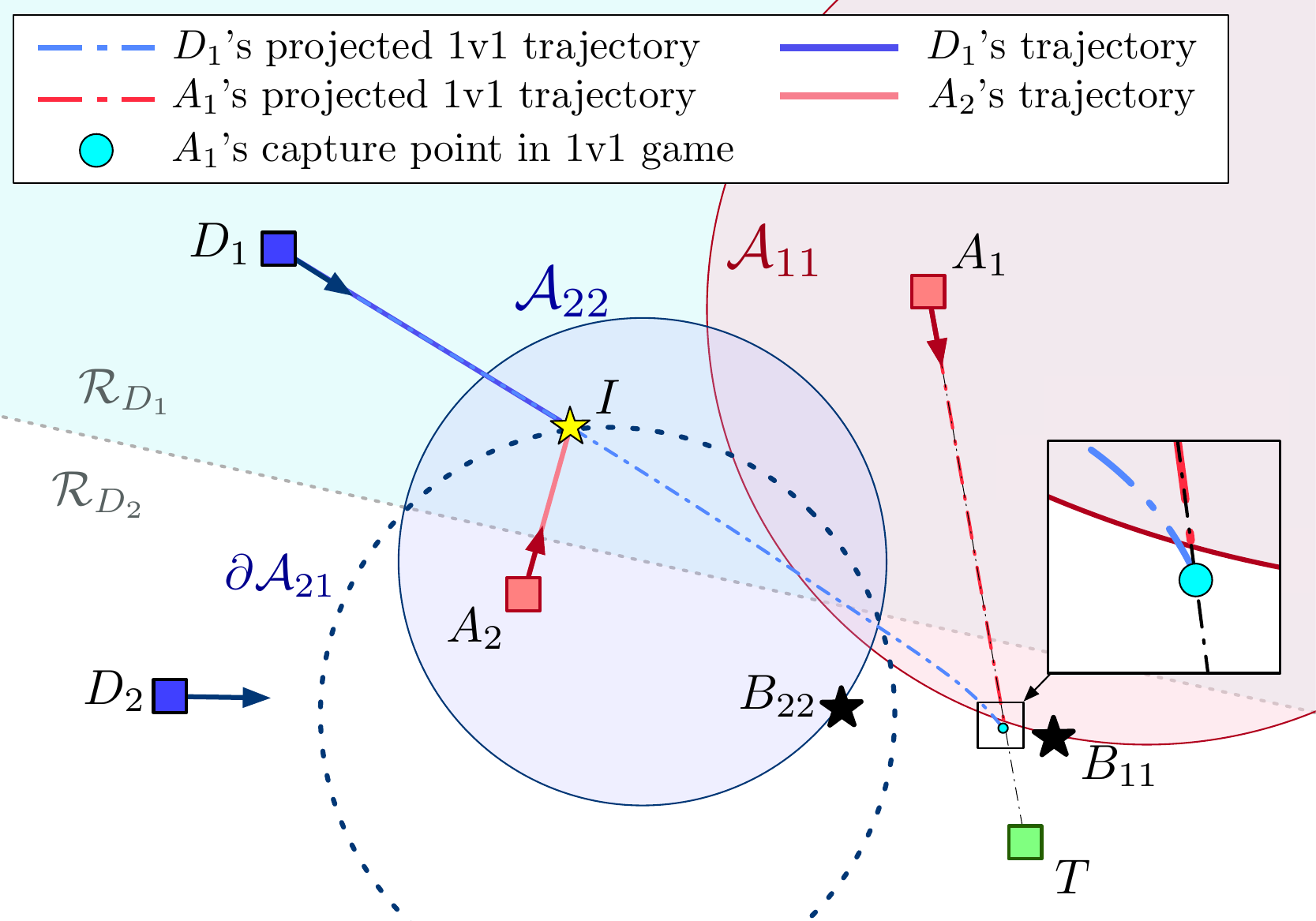}
    \caption{Illustration of two-attacker deviation scenario, where $A_1$ moves towards the target with a straight-line path, and $A_2$ seeks to intercept its pursuer $D_1$.
    }
    \label{fig: two_att_dev}
\end{figure}

\noindent When the initial conditions of the game is such that $A_2$ can assist $A_1$ by intercepting $D_1$, it naturally prompts the question: Can $A_1$  also deviate from the nominal strategy to reach the target more effectively?

As a first step, we assume that $A_1$ deviates from \eqref{eq: nom strategies A1} and simply heads straight towards the target:
\begin{align}\label{eq: A_1 dev strat}
   \gamma_{A_1}(\mbz_0) = \unitVector{\mbx_{A_1}}{\mbx_T}.
\end{align}
From Lemma~\ref{lem: J improve after intercept}, this strategy effectively ensures that the attacker team can improve their payoff
if $A_2$ can intercept $D_1$.

The problem with $A_1$ deviating from the nominal path is that the trajectory of $D_1$ becomes non-linear, as indicated by Lemma~\ref{lemma: xB_dot} and Remark~\ref{remark: straightline-condition}. Therefore, even if there exists an interception point $\mathbf{x}_I$ that satisfies \eqref{eq: intercept points xI} at the initial time, this point may not exist at $t_{f_1}$ during $D_1$'s pursuit due to its non-linear path. 
Since determining $\mathbf{x}_I$ 
requires the knowledge of the defender's position $\mathbf{x}_{D_1}$ at $t_{f_1}$ solving for the entire trajectory analytically would be too complicated. Thus, one possible approach is to numerically compute $D_1$'s trajectory at $t_0$ and solve for  $\mathbf{x}_I(t_{f_1})$. 

\begin{figure}[tbp]
\centering    \includegraphics[width=0.45\textwidth]{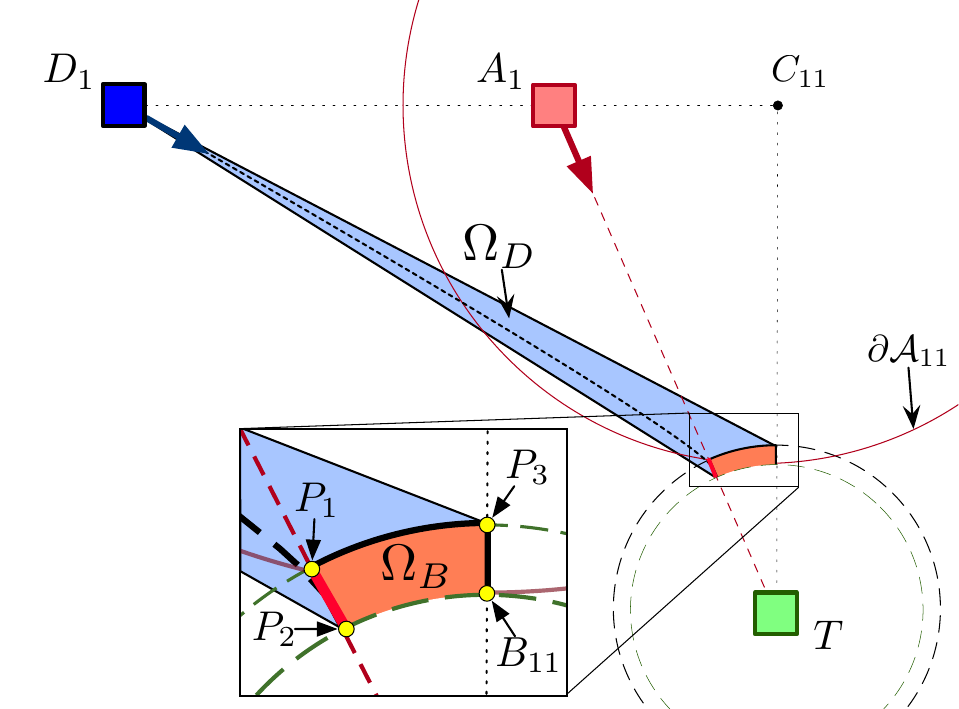}
    \caption{Illustration of points $P_1,P_2, P_3$, and bounded regions $\Omega_B$ and $\Omega_D$ for 1v1 scenario between $D_1$ and $A_1$.}
    \label{fig: bounded-regions}
\end{figure}

Now we take a geometric approach to find the sufficient conditions for $A_2$ to be able to intercept $D_1$ (see Figure~\ref{fig: bounded-regions}). Suppose $\cL(\mbx_{A_1},\mbx_T)$ intersects with $\partial\cA_{11}$ and $\partial\cB_{\rho_T}(\mbx_T)$ at $\mbx_{P_1}$ and $\mbx_{P_2}$, respectively, where  $ \rho_T(0) \triangleq \distance{\mbx_T}{\mbx_{B}} = \distance{\mbx_T}{\mbx_{P_2}}$\footnote{Since both $\mbx_B$, $\mbx_{P_2} \in \partial \ball{\rho_T}{\mbx_T}$.}  and $\cB_{\rho_T}(\mbx_T)$ denote the \emph{safe circle} around the target. 

\begin{lm}\label{lem: annular bound}
In a 1v1 scenario, if $A_1$ and $D_1$ follow state-feedback strategies given by \eqref{eq: A_1 dev strat} and \eqref{eq: gamma_D 1v1}, respectively, then $D_1$ captures $A_1$ at point $\mbx_{A_1}^{1v1}(t_f)$ whose distance from $\mbx_T$ satisfies the following conditions:
\begin{equation}\label{eq: A_1 capture gurantee}
\rho_T < \distance{\mbx_T}{\mbx_{A_1}^{1v1}(t_f)} < \distance{\mbx_T}{\mbx_{P1}}.
\end{equation}   
\end{lm}
\begin{proof}
By definition of $\cA_{11}$, $A_1$ can reach $\mbx_{P_1}$, and since $\mbx_B$ is optimal capture point which satisfies \eqref{eq: xB 1v1}, $D_1$ can guarantee $\rho_T < \distance{\mbx_T}{\mbx_{A_1}^{1v1}(t_f)}$. Thus the capture must occur satisfying  \eqref{eq: A_1 capture gurantee}.
\end{proof}
Lemma~\ref{lem: annular bound} bounds the trajectory of the capture point, $\mbx_B(t)$ inside an annular region and assures that the capture of $A_1$ happens on the line segment connecting $P_1$ and $P_2$: 
\begin{equation} \label{eq: xB_tf}
    \mbx_{B}(t_f) \in \cL(\mbx_{P_1},\mbx_{P_2}).
\end{equation}
Let $\hat{\mathbf{n}}$ denote a normal vector at $\mbx_B$ on $\cL(\mbx_{B},\mbx_{C_{11}})$, which points towards the side $A_1$ resides.
Given the strategy for $A_1$ and $D_1$ defined in \eqref{eq: A_1 dev strat} and \eqref{eq: gamma_D 1v1}, respectively, $\dot{\mathbf{x}}_B$ satisfies the following condition:
\begin{equation}\label{eq: xB_dot heading}
\hat{\mathbf{n}} \cdot \dot{\mbx}_B \geq 0.
\end{equation}

Let $H\triangleq \text{Co}\{\mathbf{x}_T,\mathbf{x}_{C_{11}},\mathbf{x}_{A_{11}}\}$ denote the convex hull of $\mathbf{x}_T$, $\mathbf{x}_{C_{11}}$, and $\mathbf{x}_{A_{11}}$,
which is a triangle on the plane. 
Following from Lemma~\ref{lem: annular bound}, let's define the annular region where the final capture would happen as follows:
\begin{align}
\Theta = \left\{ \mathbf{x} \in \mathbb{R}^2 \, \big| \, \rho_T < \|\mathbf{x} - \mathbf{x}_T\| < \rho_{P_1} \right\},
\end{align}
with $\rho_{P_1} \triangleq  \distance{\mbx_T}{\mbx_{P_1}}$.
Using Lemma~\ref{lem: annular bound}, \eqref{eq: xB_tf} and \eqref{eq: xB_dot heading}, $\mathbf{x}_B(t)$ will be bounded inside an annular-sector region $\Omega_B$ defined as follows:
\begin{equation}
\Omega_B = H \cap \Theta.    
\end{equation}

We introduce a point $P_3$ on $\partial\ball{\rho_{P_1}}{\mbx_{T}}$, such that $\unitVector{\mbx_T}{\mbx_{B}}\cdot \unitVector{\mbx_T}{\mbx_{P_3}} = \pm 1$, which suggests that $T, B,$ and $P_3$ are co-linear.
\begin{lm}\label{lem: OmegaB-OmegaD}
In a 1v1 scenario, if $\mbx_B(t)\in \Omega_B$ for $t\in [0,t_f]$, then $\mbx_{D_1}(t) \in \Omega_D \cup \Omega_B$, $t\in [0,t_f]$, where the set $\Omega_{D}$ is defined as follows:
\begin{align}
\Omega_D = 
\Omega_{D_1} \cup \Omega_{D_2},
\end{align}
with $\Omega_{D_1} \triangleq$ $\text{\emph{Co}}\{\mbx_{D_1},\mbx_{P_1},\mbx_{P_2}\}$, and $\Omega_{D_2} \triangleq$ $\emph{Co}\{\mbx_{D_1},\mbx_{P_1},\mbx_{P_3}\} \setminus \ball{\rho_{P_1}}{\mbx_T} $. 
\end{lm}
\begin{proof}
    The proof follows from the fact that $D_1$ always moves towards $\mbx_{B}(t)$, and that $\mbx_{B}(t)$ is bounded inside $\Omega_B$.
\end{proof}

Now we are ready to present the condition required for the existence of $\mbx_I$ in the following theorem.
\begin{theorem}\label{theorem: 2 - both attcker dev}
For a 2v2-TDG with dynamics described in~\eqref{eq: kinematics} and the payoff function given in~\eqref{eq: J payoff}, suppose $D_1$ and $D_2$ use the assignment-based strategies given in~\eqref{eq: nom strategies D1} and \eqref{eq: nom strategies D2}, respectively, and $A_1$ uses a deviation strategy in~\eqref{eq: A_1 dev strat}. Then $A_2$ has a strategy to intercept $D_1$ at the end of Phase I if the following condition holds:
\begin{equation} \label{eq: xI for 2 dev}
\mathcal{L}(\mathbf{x}_{D_1},\mathbf{x})\cap \mathcal{A}_{21}(0) \cap \mathcal{A}_{22}(0) \neq \varnothing, \forall \mathbf{x} \in \Omega_B.
\end{equation}
\end{theorem}

\begin{proof}
From Lemma~\ref{lem: OmegaB-OmegaD}, the defenders' trajectory is constrained within a sector defined by two lines, $\cL(\mathbf{x}_{D_1},\mathbf{x}_{P_2})$ and $\cL(\mathbf{x}_{D_1},\mathbf{x}_{P_3})$. Furthermore, according to Lemma~\ref{lem: annular bound}, the capture of $A_1$ is constrained to occur on the line segment $\cL(\mathbf{x}_{P_1},\mathbf{x}_{P_2})$ (assuming $A_2$'s absence), as the attacker remains on this line until capture. Since $\cL(\mbx_{D_1},\mbx)\cap \cA_{21}(0) \cap \cA_{22}(0) \neq \emptyset, \forall \mbx \in \Omega_B$ denotes all the points that the defenders path cross through its dominance-region\footnote{$A_2$ can reach any point inside its dominance region before both $D_1$ and $D_2$, given $A_2$ follows straight-path.}, $A_2$ has a guaranteed strategy to intercept $D_1$. 
\end{proof}

The strategy for $A_2$ is described as \eqref{eq: A2 sac gamma one dev}, where the interception point $\mbx_I$ can be found by numerically computing $D_1$'s trajectory in a 1v1 scenario against $A_1$. 

\begin{lm}
    The attacker team can win the game if \eqref{eq: xI for 2 dev} is satisfied and $\mbx_T \in \cA_{12}(0)$.
\end{lm}
\begin{proof}
If \eqref{eq: xI for 2 dev} is satisfied, $D_1$'s nominal trajectory must pass through $\cA_{21}$ and $\cA_{22}$. By Theorem~\eqref{theorem: 2 - both attcker dev}, $A_2$ has a strategy to intercept $D_1$ before getting captured by $D_2$. Therefore, if $\mbx_T \in \cA_{12}(0)$ after interception $A_1$ is capable of reaching the target and win the game.
\end{proof}
This concludes our analysis of role selection and attacker-deviation scenarios in 2v2-TDG. Next, we will present numerical examples to better illustrate the proposed strategies and outcome of the game.
\section{Examples}
\noindent In this section, we show three examples for the 2v2-TDG. The initial conditions and parameters used are as follows:
$\mbx_T = [0,0]^\top$, $\mbx_{D_1} = [-1.5, 0.7]^\top$, $\mbx_{D_2} = [-1.7, 0.3]^\top$, $\mbx_{A_1} = [-0.9, 0.7]^\top$, $\mbx_{A_2} = [-1.2, 0.4]^\top$, $\nu = 2/3$.

\subsection{Example 1 Nominal Strategies}
\begin{figure}[bp]
\centering    \includegraphics[width=0.49\textwidth]{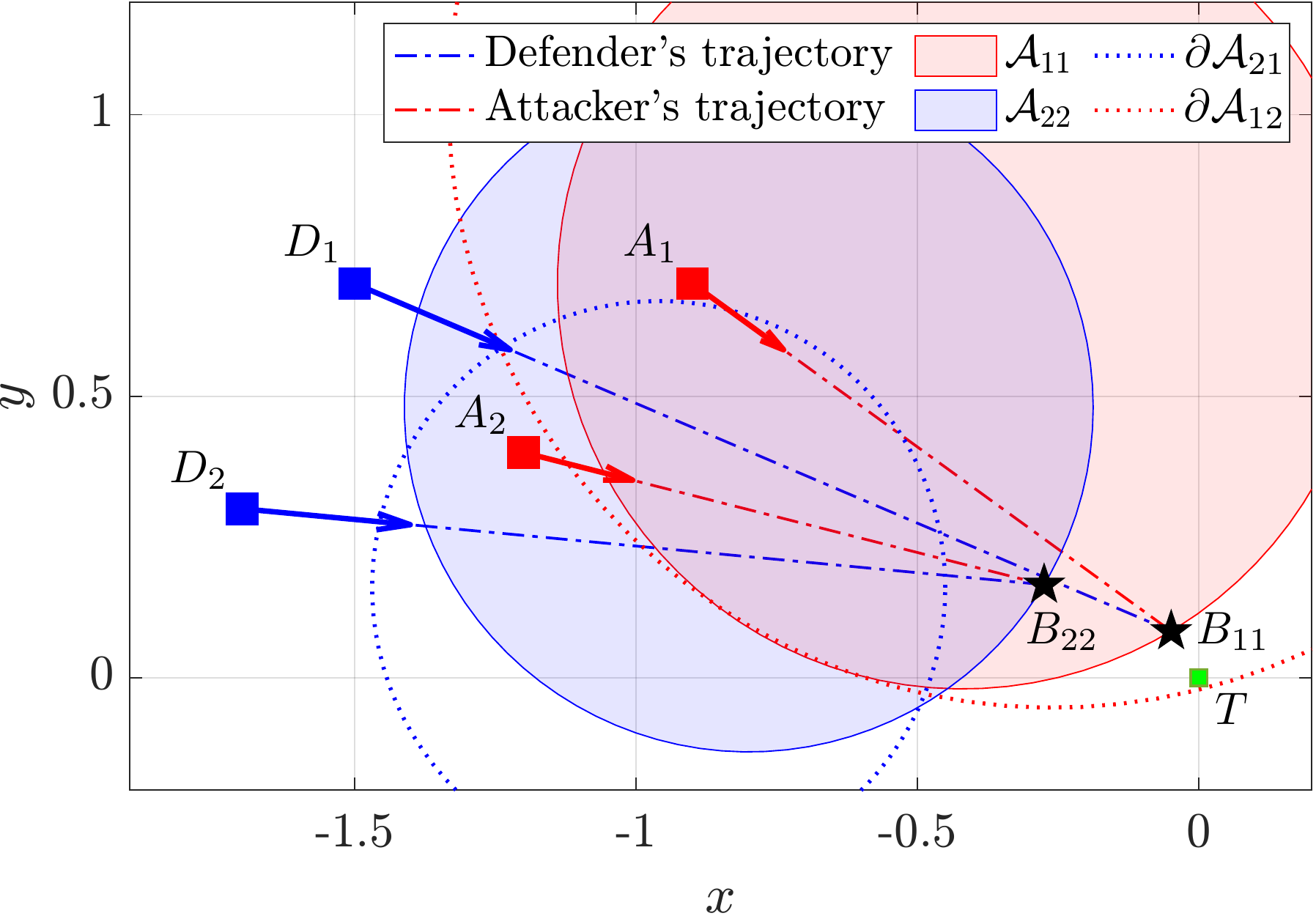}
    \caption{Nominal trajectory of the players. Defender team wins by capturing $A_1$ and $A_2$,  at $\mbx_{B_{11}}$ and $\mbx_{B_{22}}$, respectively, before they reach the target. 
    }
    \label{fig: nominal game}
\end{figure}
\noindent Figure~\ref{fig: nominal game} illustrates the nominal game scenario, showing the players' initial conditions and their nominal paths until the game terminates. From the simulation, the capture of the critical attacker $A_1$ by $D_1$ occurs at $\mathbf{x}_B = [-0.0496, 0.0826]^\top$. The payoff/value of the game is found to be $J = 0.0963 > 0$, which suggests that the game is in the defender-win region.

\subsection{Example 2 One Attacker Deviation}
\begin{figure}[tbp]
\centering    \includegraphics[width=0.49\textwidth]{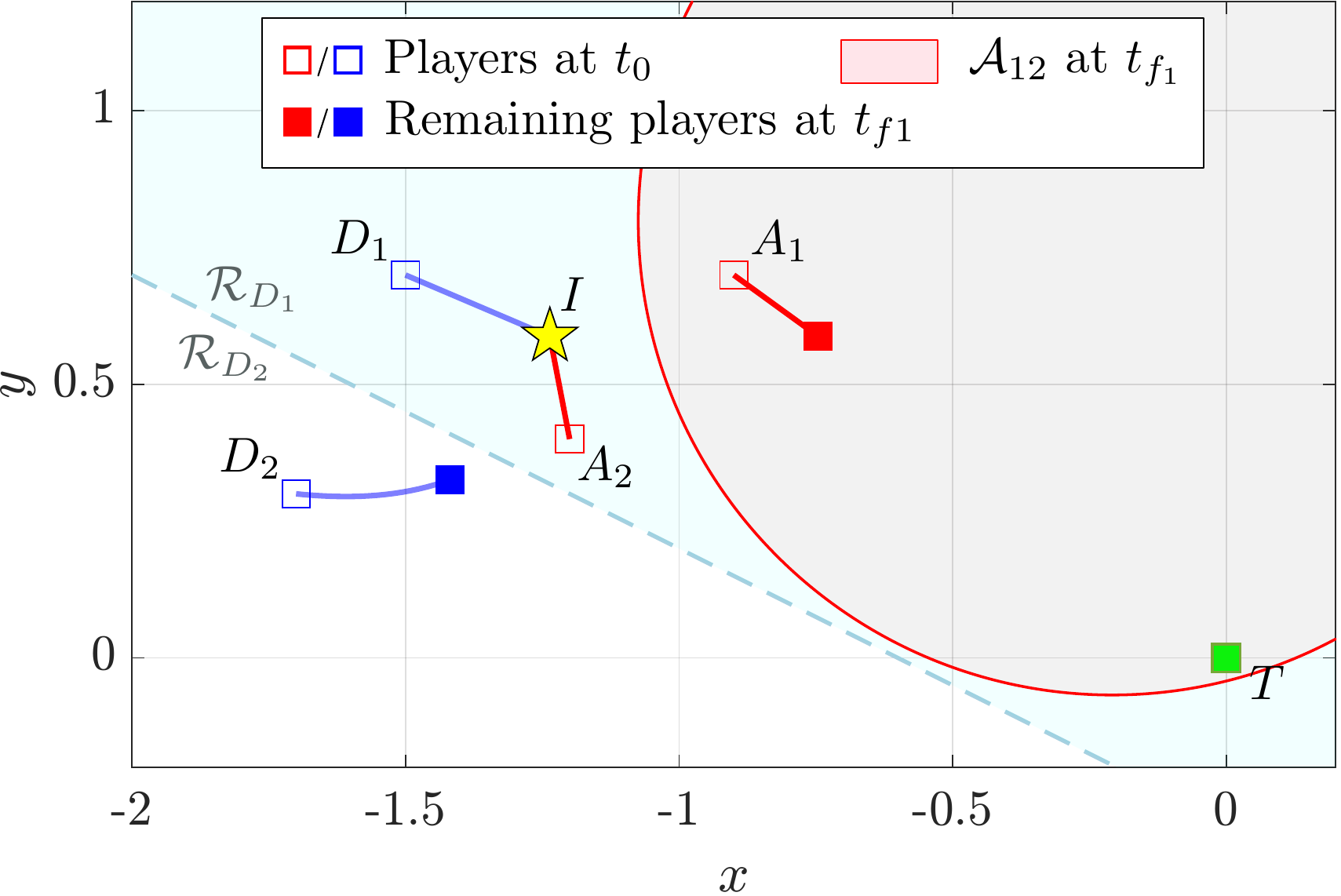}
    \caption{Game condition and the trajectories of the agents at the end of Phase I ($t_{f_1})$. $A_2$ intercepted $D_1$ and $A_1$ is capable of reaching the target.}
    \label{fig: 1 deviation}
\end{figure}
\noindent Figure~\ref{fig: 1 deviation} illustrates the end of Phase~I of the game when strategy in Corollary~\eqref{cor: 1} is used. The interception points on $\partial\mathcal{A}_{21}(0)$ are found as $\mathbf{x}_I = [-1.2362, 0.5877]^\top$ and $[-0.4603, 0.2574]^\top$. $A_2$ chooses the former one, deviates from the nominal path and intercepts $D_1$ at $t_{f_1} = 0.19$. Subsequent to $D_1$'s interception, Phase~II starts between $A_1$ and $D_2$. Note that $\mathbf{x}_T \in \mathcal{A}_{12}(t_{f_1})$, indicating that $A_1$ reaches the target before $D_2$ can reach it, thus ensuring a win for the attacker team.

\subsection{Example 3 Two Attacker Deviation}
\begin{figure}[htbp]
\centering    
\includegraphics[width=0.49\textwidth]{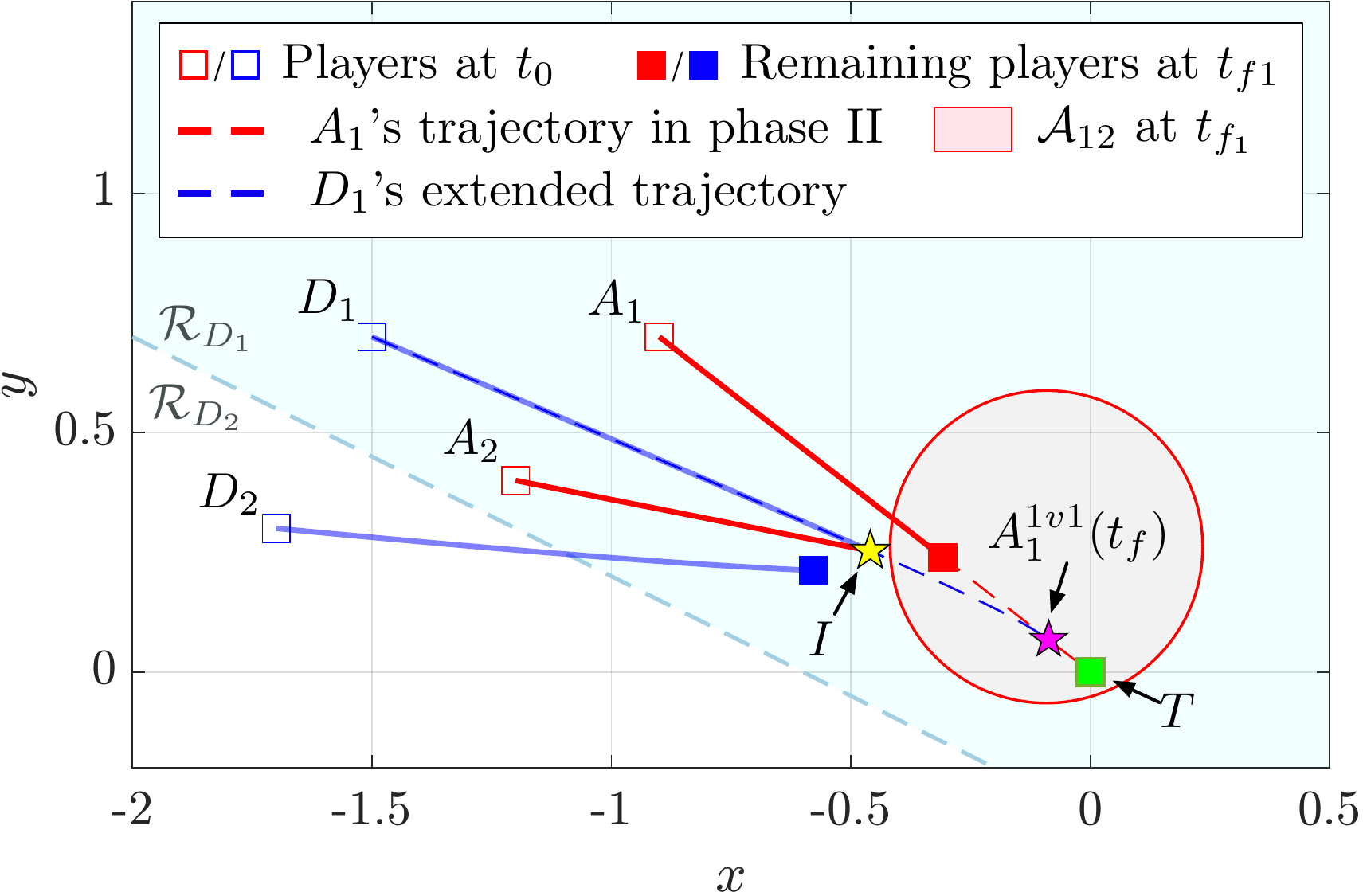}
    \caption{Trajectories of the players for two attacker deviation scenario: $A_2$ intercepts $D_1$ by anticipating the feedback-trajectory in a 1v1 scenario with $A_1$, and intercepts it at point $I$.}
    \label{fig: 2 deviation}
\end{figure}
 \noindent Figure~\ref{fig: 2 deviation} shows the game when $A_1$ and $A_2$ both deviates from the nominal path, where $A_1$ goes straight to the target and $A_2$ seeks to intercept $D_1$. The difference between nominal and computed path for the entire 1v1 game for critical attacker defender pair $A_1$-$D_1$ is shown. Utilizing the computed trajectories, $A_2$ determines the interception point as $\mbx_I = [-0.4595, 0.2530 ]^\top$ and successfully intercepts $D_1$ at $t_{f_1} = 0.75$.  Finally, in Phase~II, $\mbx_T \in \cA_{21}$, thus attacker team wins the game.
\section{Conclusion}
\noindent In this paper, we identify and present a significant gap in the traditional approach to solving team-vs-team differential games using the decomposition method, which inherently lacks the concept of team awareness due to the nature of the solutions it provides. We have shown that, in the absence of team awareness on the defender's side, the attacker team can change the outcome of the game to their favor by selecting heterogeneous roles. This paper thus motivates the pursuit of true equilibrium solutions for team-vs-team games, not only by relying on decomposition but also by focusing on the teaming aspect and choosing individual roles to outplay the opponents.

%
\bibliographystyle{IEEEtran}
\bibliography{refs}

\begin{thebibliography}{10}
\providecommand{\url}[1]{#1}
\csname url@samestyle\endcsname
\providecommand{\newblock}{\relax}
\providecommand{\bibinfo}[2]{#2}
\providecommand{\BIBentrySTDinterwordspacing}{\spaceskip=0pt\relax}
\providecommand{\BIBentryALTinterwordstretchfactor}{4}
\providecommand{\BIBentryALTinterwordspacing}{\spaceskip=\fontdimen2\font plus
\BIBentryALTinterwordstretchfactor\fontdimen3\font minus
  \fontdimen4\font\relax}
\providecommand{\BIBforeignlanguage}[2]{{%
\expandafter\ifx\csname l@#1\endcsname\relax
\typeout{** WARNING: IEEEtran.bst: No hyphenation pattern has been}%
\typeout{** loaded for the language `#1'. Using the pattern for}%
\typeout{** the default language instead.}%
\else
\language=\csname l@#1\endcsname
\fi
#2}}
\providecommand{\BIBdecl}{\relax}
\BIBdecl

\bibitem{weintraub2020introduction}
I.~E. Weintraub, M.~Pachter, and E.~Garcia, ``An introduction to
  pursuit-evasion differential games,'' \emph{2020 American Control Conference
  (ACC)}, pp. 1049--1066, 2020.

\bibitem{Issacs1965}
R.~Isaacs, \emph{Differential Games: A Mathematical Theory with Applications to
  Optimization, Control and Warfare.}\hskip 1em plus 0.5em minus 0.4em\relax
  Wiley, New York, 1965.

\bibitem{von2022circular}
A.~Von~Moll, M.~Pachter, D.~Shishika, and Z.~Fuchs, ``Circular target defense
  differential games,'' \emph{IEEE Trans. on Automatic Control}, 2022.

\bibitem{das2022guarding}
G.~Das and D.~Shishika, ``Guarding a translating line with an attached
  defender,'' \emph{2022 American Control Conference (ACC)}, pp. 4436--4442,
  2022.

\bibitem{bajaj2022competitive}
S.~Bajaj, E.~Torng, S.~D. Bopardikar, A.~Von~Moll, I.~Weintraub, E.~Garcia, and
  D.~W. Casbeer, ``Competitive perimeter defense of conical environments,''
  \emph{2022 IEEE 61st Conference on Decision and Control (CDC)}, pp.
  6586--6593, 2022.

\bibitem{lee2024Guarding}
Y.~Lee, G.~Das, D.~Shishika, and E.~Bakolas, ``Guarding a target area from a
  heterogeneous group of cooperative attackers,'' \emph{2024 American Control
  Conference (ACC)}, pp. 233--238, 2024.

\bibitem{garcia2019strategies}
E.~Garcia, A.~Von~Moll, D.~W. Casbeer, and M.~Pachter, ``Strategies for
  defending a coastline against multiple attackers,'' \emph{2019 IEEE 58th
  conference on decision and control (CDC)}, pp. 7319--7324, 2019.

\bibitem{Garcia2021_NvM}
E.~Garcia, D.~W. Casbeer, A.~Von~Moll, and M.~Pachter, ``Multiple pursuer
  multiple evader differential games,'' \emph{IEEE Trans. on Automatic
  Control}, vol.~66, Art. no.~5, pp. 2345--2350, 2021.

\bibitem{bakolas2010optimal}
E.~Bakolas and P.~Tsiotras, ``Optimal pursuit of moving targets using dynamic
  voronoi diagrams,'' \emph{49th IEEE Conf. on decision and control (CDC)}, pp.
  7431--7436, 2010.

\bibitem{pachter2022strategies}
M.~Pachter, D.~W. Casbeer, and E.~Garcia, ``Strategies for target defense from
  a fast attacker,'' \emph{2022 IEEE Conference on Control Technology and
  Applications (CCTA)}, pp. 233--238, 2022.

\bibitem{dorothy2024one}
M.~Dorothy, D.~Maity, D.~Shishika, and A.~Von~Moll, ``One apollonius circle is
  enough for many pursuit-evasion games,'' \emph{Automatica}, vol. 163, Art.
  no. 111587, 2024.

\bibitem{goutam2024defending}
G.~Das, M.~Dorothy, Z.~I. Bell, and D.~Shishika, ``Defending a static target
  point with a slow defender,'' \emph{2024 American Control Conference (ACC)},
  pp. 4064--4071, 2024.

\bibitem{Kuhn1955Hungarian}
H.~W. Kuhn, ``{The Hungarian Method for the Assignment Problem},'' \emph{Naval
  Research Logistics Quarterly}, vol.~2, Art. no. 1--2, pp. 83--97, March 1955.

\bibitem{Shishika2018}
D.~Shishika and V.~Kumar, ``Local-game decomposition for multiplayer
  perimeter-defense problem,'' \emph{2018 IEEE Conf. on Decision and Control
  (CDC)}, 2018.

\bibitem{sun2019multiplayer}
W.~Sun, P.~Tsiotras, and A.~J. Yezzi, ``Multiplayer pursuit-evasion games in
  three-dimensional flow fields,'' \emph{Dynamic Games and Applications},
  vol.~9, pp. 1188--1207, 2019.

\bibitem{yan2019task}
R.~Yan, Z.~Shi, and Y.~Zhong, ``Task assignment for multiplayer reach--avoid
  games in convex domains via analytical barriers,'' \emph{IEEE Trans. on
  Robotics}, vol.~36, Art. no.~1, pp. 107--124, 2019.

\bibitem{bakolas2021decentralized}
E.~Bakolas and Y.~Lee, ``Decentralized game-theoretic control for dynamic task
  allocation problems for multi-agent systems,'' \emph{2021 American Control
  Conf. (ACC)}, pp. 3228--3233, 2021.

\bibitem{zhang2021pursuer}
L.~Zhang, A.~Prorok, and S.~Bhattacharya, ``Pursuer assignment and control
  strategies in multi-agent pursuit-evasion under uncertainties,''
  \emph{Frontiers in Robotics and AI}, vol.~8, Art. no. 691637, 2021.

\bibitem{mittal2018pursuit}
A.~Mittal, A.~Jain, A.~Kumar, and R.~Tiwari, ``Pursuit-evasion: Multiple
  pursuer pursue multiple evader using wavefront and hungarian method,''
  \emph{Proceedings of the International Conf. on Computing and Communication
  Systems: I3CS 2016, NEHU, Shillong, India}, pp. 473--488, 2018.

\bibitem{burkard2012assignment}
R.~Burkard, M.~Dell'Amico, and S.~Martello, \emph{Assignment problems: revised
  reprint}.\hskip 1em plus 0.5em minus 0.4em\relax SIAM, 2012, pp. 172--173.

\end{thebibliography}

\end{document}